\documentclass[12pt]{article}

\usepackage{indentfirst}
\usepackage{graphicx}

\usepackage{amsmath}

\begin{document}
\thispagestyle{empty}
\newcommand{\unmedio}{{\scriptstyle\frac{1}{2}}}
\newcommand{\eff}{_{\text{eff}}}
\newcommand{\Infinity}{\infty}
\newcommand{\flip}{_{\text{flip}}}
\newcommand{\bos}{_{0\text{B}}}
\newcommand{\bosonic}{_{\text{bos}}}
\newcommand{\ferm}{_{0\text{F}}}
\newcommand{\trial}{_{\text{trial}}}
\newcommand{\true}{_{\text{true}}}
\newcommand{\Ci}{\operatorname{Ci}}
\newcommand{\tr}{\operatorname{tr}}

\newcommand{\PsiB}{\bar{\Psi}}   
\newcommand{\phiH}{\hat{\phi}}
\newcommand{\etaH}{\hat{\eta}}
\newcommand{\chiB}{\bar{\chi}}
\newcommand{\xiH}{\hat{\xi}}
\newcommand{\zetaH}{\hat{\zeta}}
\newcommand{\vH}{\hat{v}}
\newcommand{\bH}{\hat{b}}

\newcommand{\AB}{\bar{A}_\mu}
\newcommand{\BB}{\bar{B}_\mu}
\newcommand{\AT}{\tilde{A}_\mu}
\newcommand{\BT}{\tilde{B}_\mu}

\newcommand{\slp}{\raise.15ex\hbox{$/$}\kern-.57em\hbox{$\partial$}}
\newcommand{\slA}{\raise.15ex\hbox{$/$}\kern-.63em\hbox{$A$}}

\newcommand{\difp}{\frac{d^2p}{(2\pi)^2}\,}

\newcommand{\bra}{\left\langle}
\newcommand{\ket}{\right\rangle}
\newcommand{\bracket}{\left\langle\,\right\rangle}

\newcommand{\D}{\mathcal{D}}
\newcommand{\N}{\mathcal{N}}
\newcommand{\Lag}{\mathcal{L}}
\newcommand{\V}{\mathcal{V}}
\newcommand{\Z}{\mathcal{Z}}

\title{Conformal Gaussian Approximation}
\author{An\'{\i}bal Iucci$^a$ and Carlos M. Na\'on$^a$}
\date{\today}

\maketitle
\begin{abstract}
We present an alternative way to determine the unknown parameter associated to a
gaussian approximation in a generic two-dimensional model. Instead of the standard
variational approach, we propose a procedure based on a quantitative prediction of
conformal invariance, valid for systems in the scaling regime, away from criticality.
We illustrate our idea by considering, as an example, the sine-Gordon model. Our
method gives a good approximation for the soliton mass as function of $\beta$.
\end{abstract}
\vspace{3cm} Pacs: 11.10.Lm\\ \hspace*{1,7 cm} 11.10.Kk

\noindent --------------------------------

\noindent $^a$ {\footnotesize Instituto de F\'{\i}sica La Plata,
Departamento de F\'{\i}sica, Facultad de Ciencias Exactas,
Universidad Nacional de La Plata.  CC 67, 1900 La Plata,
Argentina.\\emails: iucci@fisica.unlp.edu.ar,
naon@fisica.unlp.edu.ar}

\newpage

The so called ``self consistent harmonic approximation" (SCHA) is
a non-perturbative technique that has been extensively employed in
Statistical Mechanics \cite{Saito}\cite{Fisher-Zwerger} and
Condensed Matter physics
\cite{Gogolin}\cite{Egger}\cite{Xu}\cite{Iucci} applications.
Roughly speaking it amounts to replacing an exact action
$S_{\text{true}}$ by a trial action $S_{\text{trial}}$ that makes
the problem tractable. Usually $S_{\text{trial}}$ is just a
quadratic action that depends on certain unknown parameter
$\Omega$ that must be determined through some criterion such as
the minimization of the free energy of the system. This
approximation is intimately related the the ``gaussian effective
potential" (GEP) \cite{Stevenson}\cite{Ingermanson} in Quantum
Field Theories (QFT's), a variational approximation to the
effective potential which uses a gaussian wave functional
depending on some mass parameter as the trial ground state. It
also relies on a minimization principle often called ``principle
of minimal sensitivity" \cite{minimal} to determine the additional
parameter. The main purpose of this Letter is to point out that in
two-dimensional problems (1+1 QFT's) there is an alternative way
to obtain the quantity $\Omega$. Since, as we shall see, this
method is based on Conformal Field Theory (CFT) \cite{CFT}, we
call it ``conformal gaussian approximation" (CGA). In view of the
interesting physics that is currently under investigation in
low-dimensional systems (organic conductors \cite{organic
conductors}, charge transfer salts \cite{salts}, quantum wires
\cite{quantum wires}, edge states in a two-dimensional (2D)
electron system in the fractional quantum Hall regime \cite{FQH},
Carbon Nanotubes (CNT) \cite{CNT}, Luttinger-like systems in 2D
high temperature superconductors \cite{Orgad}) the dimensional
restriction in the validity of our approach does not diminish its
practical interest.

We shall begin by depicting the main features of the standard SCHA. One starts from a
partition function

\begin{equation}
\Z\true=\int\D\mu e^{-S_{\text{true}}}
\end{equation}

\noindent where $\D\mu$ is a generic integration measure. An elementary manipulation
leads to

\begin{equation}\label{eq:Z}
\Z_{\text{true}}=\frac{\int\D\mu
e^{-(S_{\text{true}}-S_{\text{trial}})}\,e^{-S_{\text{trial}}}}{\int\D\mu
e^{-S_{\text{trial}}}}\int\D\mu e^{-S_{\text{trial}}}=\Z_{\text{trial}}\bra
e^{-(S_{\text{true}}-S_{\text{trial}})} \ket_{\text{trial}}
\end{equation}

\noindent for any trial action $S_{\text{trial}}$. Now, by means of the property

\begin{equation}
\bra e^{-f} \ket\geq e^{-\bra f \ket},
\end{equation}

\noindent for $f$ real, and taking natural logarithm in equation
(\ref{eq:Z}), we obtain Feynman's inequality \cite{Feynman}

\begin{equation}\label{Feynman}
\ln\Z\true\geq \ln\Z\trial - \bra S\true-S\trial\ \ket\trial.
\end{equation}
In general, $S\trial$ depends on some parameters, which are fixed
by minimizing the right hand side of the last equation.

From now on, in order to illustrate the procedure, we shall consider scalar 2D models,
with action

\begin{equation}
S\true=\int\difp\varphi(p)\frac{F(p)}{2}\varphi(-p) + \int d^2x\,\
V(\varphi)
\end{equation}
where $\varphi(p)$ is a scalar field and $F(p)$ is usually of the form $F(p)\sim p^2$.
For simplicity, in this formula we have written the kinetic term in Fourier space but
we kept the interaction term in coordinate space.

As the trial action one proposes a quadratic one,

\begin{equation}
S\trial=\int\difp\left[\varphi(p)\frac{F(p)}{2}\varphi(-p) +
\frac{\Omega^2}{2}\varphi(p)\varphi(-p)\right],
\end{equation}
where $\Omega$ is the trial parameter. To proceed with the
minimization, we first write the partition function for the trial
action in the form

\begin{equation}\label{ztrial}
\Z\trial=\exp{\left[-\frac{\V}{2} I_0(\Omega)\right]}
\end{equation}

\noindent where $\V$ is the volume (infinite) of the whole space
$\int d^2x$, and we have defined

\begin{equation}
I_0(\Omega)=\int \frac{d^2p}{(2\pi)^2} \ln[F(p)+\Omega^2].
\end{equation}

\begin{equation}
I_n(\Omega)=\int \frac{d^2p}{(2\pi)^2} \frac{1}{[F(p)+\Omega^2]^n}
\end{equation}
with the formal properties

\begin{equation}
\frac{dI_0(\Omega)}{d\Omega}=2\Omega I_1(\Omega)
\end{equation}

\begin{equation}
\frac{dI_n(\Omega)}{d\Omega}=-2\Omega n I_{n+1}(\Omega).
\end{equation}

To go further we must specify the potential $V(\varphi)$. Let us specialize the
discussion to the sine-Gordon model (SGM), taking

\begin{equation}
V(\varphi)=\frac{\alpha}{\beta^2}\left[1-\cos(\beta\varphi)\right],
\end{equation}
and $F(p)=p^2$.

As it stands the theory is divergent. To take into account the divergences we can
implement Coleman's normal order prescription \cite{Coleman} with respect to any given
mass-dimensional constant $\rho$ from the beginning \cite{beg}, by replacing in the
lagrangian $V(\varphi)$ by its normal ordered form

\begin{equation}
\N_\rho[V(\varphi)]=\frac{\alpha}{\beta^2}\left[1-\cos(\beta\varphi)e^{\unmedio\beta^2
I_1(\rho)}\right]
\end{equation}
where $\N_\rho[...]$ means the normal ordering form with respect
to $\rho$.

It is now straightforward to compute $\langle S\true-S\trial
\rangle$, by following, for instance, the steps explained in ref.
\cite{Li-Naon}. The result is

\begin{equation}\label{elegant}
\bra S\true-S\trial \ket\trial =
\V\left[\frac{\alpha}{\beta^2}\left(1-e^{-\unmedio\beta^2[I_1(\Omega)-I_1(\rho)]}\right)
-\frac{\Omega^2}{2}I_1(\Omega)\right].
\end{equation}

\noindent Now inserting equations (\ref{ztrial}) and
(\ref{elegant}) in equation (\ref{Feynman}), and extremizing the
r.h.s. with respect to $\Omega$, we finally obtain

\begin{equation}\label{gapscha}
\Omega^2-\alpha e^{-\beta^2/2 (I_1(\Omega)-I_1(\rho))}=0.
\end{equation}
This gap equation allows to extract a finite answer for $\Omega$, depending on the
mass parameter $\rho$ (the difference $I_1(\Omega)-I_1(\rho)$ is finite). Note that
the value of $\rho$ is completely arbitrary, if one chooses it to be equal to the
trial mass $\Omega$, the solution to the equation is

\begin{equation}\label{SGGaussian}
\Omega^2=\alpha .
\end{equation}
The same result would be obtained if instead of $\rho=\Omega$ one had taken
$\rho=\sqrt{\alpha}$.

\vspace{.5cm}

Let us now present an alternative route to determine $\Omega$. To this end we will
exploit a quantitative prediction of conformal invariance for 2D systems in the
scaling regime, away from the critical point. Indeed, starting from the so called
'c-theorem' \cite{Zamolodchikov} Cardy \cite{Cardy} showed that the value of the
conformal anomaly $c$, which characterizes the model at the critical point, and the
second moment of the energy-density correlator in the scaling regime of the
non-critical theory are related by

\begin{equation} \label{Cardy}
\int d^2x\, |x|^2\, \bra\varepsilon(x)\varepsilon(0)\ket = \frac{c}{3\, \pi
\,t^2\,(2 - \Delta_\varepsilon)^2},
\end{equation}
where $\varepsilon$ is the energy-density operator, $\Delta_\varepsilon$ is its
scaling dimension and $t\propto(T-T_c)$ is the coupling constant of the interaction
term that takes the system away from criticality. The validity of this formula has
been explicitly verified for several models \cite{Cardy} \cite{Cardy2} \cite{Naon}
\cite{Salvay}. For the SGM, the energy density operator is given by the cosine term,
its conformal dimension is $\Delta_\varepsilon=\beta^2/4\pi$, $t$ is the coupling
constant $\alpha/\beta^2$ and the associated free bosonic CFT has $c=1$.

Now we claim that $\Omega$ can be determined in a completely different, not
variational way, by enforcing the validity of the above conformal identity for the
trial action. In other words, we will demand that the following equation holds:

\begin{equation} \label{Main}
\frac{\alpha^2}{\beta^4}\int d^2x\, |x|^2\, \bra
\cos\beta\varphi(x)\,\cos\beta\varphi(0) \ket\trial = \frac{1}{3\,
\pi \,\,(2 - \frac{\beta^2}{4\pi})^2},
\end{equation}

\noindent which is to be viewed as an equation for the mass parameter $\Omega$. Of
course, if one is interested in comparing the answer given by this formula with the
SCHA result, when evaluating the left hand side of (\ref{Main}) one must adopt a
regularizing prescription equivalent to the normal ordering implemented in the SCHA
calculation. A careful computation leads to the following gap equation:

\begin{equation}\label{gapcga}
(\frac{\Omega}{\rho})^{2(2-u)}=(\frac{\alpha}{\rho^2})^2\,\frac{3}{32}\,\frac{2-u}{u^2}
\end{equation}

\noindent where we have defined the variable $u=\beta^2/4\pi$
($0\leq u <2$) and $\rho$ is the normal ordering parameter, as
before. We see that, as in the standard SCHA equation
(\ref{gapscha}), one has different answers for different choices
of $\rho$, but in this case, the results obtained for the values
$\sqrt{\alpha}$ and $\Omega$ are different. In any case one gets a
non trivial dependence of $\Omega$ on $\beta^2$ in contrast with
the SCHA. This is interesting if one recalls the physical meaning
of mass gaps in the context of the SGM. Indeed, as it is
well-known, Dashen, Hasslacher and Neveu (DHN) \cite{DHN} have
computed by semiclassical techniques the mass spectrum for the
SGM. It consists of a soliton (associated to the fermion of the
Thirring model) with mass

\begin{equation}\label{soliton}
M_{sol}=\frac{2-u}{\pi\,u}\,\sqrt{\alpha},
\end{equation}

\noindent and a sequence of doublet bound states with masses

\begin{equation}\label{doublets}
M_{N}=\frac{2(2-u)}{\pi\,u}\,\sin\left[N
\frac{\pi\,u}{2(2-u)}\right]\, \sqrt{\alpha},
\end{equation}

\noindent with $N=1,2,...<(2-u)/u$. (From this last condition it is easy to see that
in order to have $N$ bound states one must have $u<2/(N+1)$. As a consequence there is
no bound state for $u>1$). Therefore the masses in the SGM spectrum also depend on $u$
as the CGA mass of equation (\ref{gapcga}). Thus, in this respect our proposal seems
to be able to improve the standard gaussian prediction for the SGM, at least
qualitatively. In order to perform a more specific and quantitative  discussion let us
compare equations (\ref{gapcga}) and (\ref{soliton}) as functions of $u$. We set
$\rho=\sqrt{\alpha}$, which corresponds to the prescription employed by DHN when
deriving (\ref{soliton}) and (\ref{doublets}). The result is shown in Fig. 1 where one
can observe a general qualitative analogy between both curves. In particular, for
$0.7\leq u \leq 1$ ($u=1$ corresponds to the free fermion point of the Thirring model
and to the Luther-Emery point in the backscattering model \cite{Luther-Emery}) our CGA
prediction is in full agreement with the values of the soliton mass as computed by
DHN. We want to stress that for $u=1$ we get
$\Omega/\sqrt{\alpha}=\sqrt{3/32}\approx0.30$ whereas the exact value given by
(\ref{soliton}) is $1/\pi\approx0.31$ (standard SCHA yields, of course,
$\Omega/\sqrt{\alpha}=1$).


\vspace{1cm} To conclude, we have reconsidered the well-known SCHA method in which a
comparatively complex action is replaced by a simpler quadratic system depending on a
mass parameter $\Omega$ which is usually determined through a variational calculation.
Taking into account the (1+1)-dimensional case, we have proposed an alternative way
for evaluating $\Omega$. Our technique (CGA) is based on a consequence of
Zamolodchikov's \cite{Zamolodchikov} c-theorem first derived by Cardy \cite{Cardy}. We
have illustrated the proposal by considering the sine-Gordon (SGM) model. We showed
that for this model CGA gives a quite good prediction for the behavior of the soliton
mass as function of $\beta^2$ (see equations (\ref{gapcga}) and (\ref{soliton}) and
Fig. 1. It would be interesting to test our approach in other models such as the
continuum version of the tricritical Ising model, which is described by the second
model of the unitary minimal series \cite{BPZ} \cite{FQS} with central charge
$c=7/10$. \vspace{0.5cm}

{\bf Acknowledgements}

This work was supported by the Consejo Nacional de Investigaciones Cient\'{\i}ficas y
T\'ecnicas (CONICET) and Universidad Nacional de La Plata (UNLP), Argentina.

\newpage

{\bf Figure caption}\\

Figure 1: Masses in units of $\sqrt{\alpha}$ as functions of $u$. The dashed line is
$M_{sol}/\sqrt{\alpha}$, whereas the filled line represents $\Omega/\sqrt{\alpha}$ as
given by CGA.\\
\end{document}